\documentstyle[epsf,12pt]{article}

\def\beeq{\begin{eqnarray}} \def\eeeq{\end{eqnarray}}
\newcommand\mysection{\setcounter{equation}{0}\section}
\renewcommand{\theequation}{\thesection.\arabic{equation}}
\newcounter{hran} \renewcommand{\thehran}{\thesection.\arabic{hran}}

\def\bmini{\setcounter{hran}{\value{equation}}
  \refstepcounter{hran}\setcounter{equation}{0}
  \renewcommand{\theequation}{\thehran\alph{equation}}\begin{eqnarray}}

\def\bminiG#1{\setcounter{hran}{\value{equation}}
\refstepcounter{hran}\setcounter{equation}{-1}
\renewcommand{\theequation}{\thehran\alph{equation}}
\refstepcounter{equation}\label{#1}\begin{eqnarray}}

\def\emini{\end{eqnarray}\relax\setcounter{equation}{\value{hran}}\renewcommand{\theequation}{\thesection.\arabic{equation}}}

\def\ben{\begin{enumerate}}  \def\een{\end{enumerate}}
\def\bit{\begin{itemize}}    \def\eit{\end{itemize}}
\def\beq{\begin{equation}}   \def\eeq{\end{equation}}
\def\bea{\begin{eqnarray}}  \def\eea{\end{eqnarray}}
\def\nn{\nonumber}
\def\noi{\noindent}
\def\sq{\hbox {\rlap{$\sqcap$}$\sqcup$}}
\def\lsim{\raise0.3ex\hbox{$<$\kern-0.75em\raise-1.1ex\hbox{$\sim$}}}
\def\gsim{\raise0.3ex\hbox{$>$\kern-0.75em\raise-1.1ex\hbox{$\sim$}}}
 \def\cite#1{[\ref{#1}]}
 \def\citd#1#2{[\ref{#1},\ref{#2}]}
 \def\citt#1#2#3{[\ref{#1},\ref{#2},\ref{#3}]}
 \def\citm#1#2{[\ref{#1}--\ref{#2}]}

\begin{document}
\vbox to 1 truecm {}
\begin{center}
{\bf MONOPOLE CONDENSATION AND ANTISYMMETRIC TENSOR FIELDS: COMPACT 
QED AND THE
WILSONIAN RG FLOW IN YANG-MILLS THEORIES} \\ \vspace{1 truecm}
{\bf Ulrich Ellwanger}\footnote{email : ellwange@qcd.th.u-psud.fr}\\
Laboratoire de Physique Th\'eorique\footnote{Unit\'e Mixte de Recherche
UMR 8627 - CNRS 
}\\    Universit\'e de Paris XI, B\^atiment 210, 91405 Orsay Cedex,
France  \end{center} \vspace{2 truecm} 
\begin{abstract}
A field theoretic description of monopole condensation in strongly
coupled gauge 
theories is given by actions involving antisymmetric tensors $B_{\mu
\nu}$ of rank 2. We 
rederive the corresponding action for $4d$ compact QED, summing
explicitly over all 
possible monopole configurations. Its gauge symmetries and Ward 
iden\-tities are
discussed. Then we consider the Wilsonian RGs for Yang-Mills theories in
the presence of 
collective fields (again tensors $B_{\mu \nu}$) for the field strengths
$F_{\mu \nu}^a$ 
associated to the U(1) subgroups. We show that a ``vector-like'' Ward
identity for the 
Wilsonian action involving $B_{\mu \nu}$, whose validity cor\-res\-ponds
to monopole 
condensation, constitutes a fixed point of the Wilsonian RG flow.
  \end{abstract}

\vspace{2 truecm} 

\noi LPT Orsay 99-24 \\ 
\noi June 1999 \\
 
\newpage
\pagestyle{plain}
\mysection{Introduction}
\hspace*{\parindent} Recently progress has been made in the field
theoretic formulation 
of monopole condensation in strongly coupled gauge theories, which has
been proposed by 
t'Hooft and Mandelstam \cite{1r} as the underlying mechanism of
confinement in QCD. 
Quevedo and Trugenberger \cite{2r} have proposed the ``kinetic parts''
(quadratic in 
the fields) of actions in $d$ space-time dimensions describing the
condensation of $(d - 
r - 1)$ dimensional topological defects by the means of antisymmetric
tensors of rank 
$r$. Applied to monopole condensation in gauge theories in $d = 4$ the
corresponding 
action requires the introduction of a Kalb-Ramond field $B_{\mu \nu}$ of
rank 2. \par 

Starting with the dual formulation of compact QED in $d = 3$ \cite{3r}
(where the 
topological defects have dimension 0), Polyakov \cite{4r} derived a
partition function 
involving $B_{\mu \nu}$, which couples to the surface of the Wilson
loop. (This coupling 
has also been considered in \cite{2r}.) One of the aims of Polyakov was
to show how a 
second quantized string theory emerges due to the multivaluedness of the
action of 
$B_{\mu \nu}$ in Minkowski space. The massive Kalb-Ramond field $B_{\mu
\nu}$ in 
Polyakov's action plays exactly the role assigned to it in
\cite{2r}. Employing a 
duality transformation beyond the semiclassical approximation (which was
used by 
Polyakov), and working in $d = 4$, Diamantini, Quevedo and Trugenberger
\cite{5r} 
rederived the result of Polyakov, again with the dual formulation of
compact QED 
(involving a massive vector in $d = 4$) as a starting point. \par

In \cite{6r} (see also \cite{7r}) we had introduced rank 2 tensor fields
in Yang-Mills 
theories as ``collective fields'' for the field strength tensor $F_{\mu
\nu}^a$, in some 
analogy with the field strength formulation of Yang-Mills theories
\citd{8r}{9r}. Using 
the Wilsonian exact renormalization group equations (ERGEs) in the
presence of the 
collective fields we have argued in \cite{6r} that the quadratic part of
the Wilsonian 
effective action in the infrared limit assumes a particular form, which
is equivalent to 
the Quevedo-Trugenberger and Polyakov action \citd{2r}{4r}. A non-trivial
``phenomenological'' test of this approach consists in the computation
of the field 
strength two-point function, which is now given in terms of the
two-point function of 
$B_{\mu\nu}$ \citd{10r}{11r}, and which agrees well with lattice
results. \par 

The aim of the present paper is twofold: First, in section 2, we reconsider
four-dimensional compact QED on the lattice. Without passing by the dual
formulation 
involving the massive vector field, we will show directly, how monopole
condensation lets 
the Kalb-Ramond field $B_{\mu\nu}$ appear, and we will rederive its
action by explicit 
summation over monopole configurations. Our approach also allows to
discuss 
ex\-pli\-ci\-tly, how vector-like gauge symmetries (under which $B_{\mu
\nu}$ transforms) 
together with gauge fixing terms appear in a formulation, where both
$B_{\mu \nu}$ and 
the original gauge field $A_{\mu}$ are present in the action. We
emphasize the role of an 
associated Ward identity in this formulation. We also discuss how
$A_{\mu}$ can be 
``gauged away'' without modifying the number of degrees of freedom,
whereupon one 
recovers Polyakov's formulation involving just a massive $B_{\mu \nu}$
field without 
manifest gauge invariance. \par

Second, in section 3, we consider the Wilsonian ERGEs for Yang-Mills
theories in the 
maximal Abelian gauge, and in the presence of collective fields $B_{\mu
\nu}^a$ for 
the ``diagonal'' components of the field strength tensor $F_{\mu \nu}^a$
(the index 
$a$ being associated with the generators of the $N-1$ U(1) subgroups of
SU(N)). We 
present a modified vector-like Ward identity (depending explicitly on
the Wilsonian 
infrared cutoff $k$) which a) is invariant under the ERG flow, and
describes thus 
fixed ``points'' (ac\-tual\-ly still an infinite dimensional stable
subclass) of Wilsonian 
effective actions, and b) which turns into the Ward identity satisfied
by the 
Quevedo-Trugenberger and Polyakov actions for $k \to 0$. The role of
this Ward 
identity for Wilsonian Yang-Mills effective action in the infrared limit
is discussed in section 4.   
 
\mysection{Compact QED}
\hspace*{\parindent} Let us start with the partition function of compact
QED on the 
lattice, following closely the presentation of Polyakov \cite{3r}. We
will work in $d = 
4$ Euclidean dimensions; our results can, however, straightforwardly be
carried over 
to arbitrary dimensions, and sometimes we will let $d$ to be arbitrary.
\par 

On a lattice with lattice spacing $\ell = 1$ the action of compact QED
is given by  
\beq
\label{2.1}
S = {1 \over 2e^2} \sum_x \left ( 1 - \cos F_{\mu \nu} (x) \right ) 
\quad ,  \eeq

\noi where the sum over $\mu$, $\nu$ at each lattice site $x$ is
understood, and 
\beq
\label{2.2}
F_{\mu \nu} = \partial_{\mu} A_{\nu} - \partial_{\nu} A_{\mu}
  \eeq 

\noi with $\partial_{\mu}$ being the lattice (forward) derivative. The
fields $A_{\mu}$ 
are restricted to the domain $- \pi \leq A_{\mu} \leq \pi$. Neglecting
higher powers 
than $F_{\mu \nu}^2$ in the action, but respecting the periodicity of
the cosine, the 
partition function can be written as \cite{3r}
\beq
\label{2.3}
Z = \sum_{n_{\mu \nu}(x)} \int_{- \pi}^{+ \pi} \prod_x \left [ dA_{\mu}
\ \delta 
(\partial_{\mu} A_{\mu}) \right ] \ e^{-{1 \over 4e^2} \sum\limits_x
(F_{\mu \nu}(x) - 
2 \pi n_{\mu \nu}(x))^2} \quad .  
  \eeq

Here the antisymmetric tensor $n_{\mu \nu}(x)$ represents a set of 6
independent 
integers (in $d= 4$) at each lattice site. For later convenience we have
added a gauge 
fixing $\delta$-function $\delta (\partial_{\mu}A_{\mu})$. Due to the
restricted domain 
of integration over $A_{\mu}$ such a gauge fixing is actually not
mandatory; it just 
amounts to the multiplication of $Z$ by a finite factor per lattice
site. \par 

Next we introduce a Hodge decomposition of the 6 independent integers 
$n_{\mu \nu}$ per lattice site:
\beq
\label{2.4}
n_{\mu \nu} = \partial_{[ \mu} m_{\nu ]} + B_{\mu \nu} \quad .  
  \eeq

\noindent The vector $m_{\mu}$ satisfies $\partial_{\mu} m_{\mu} = 0$
and represents 
thus $d - 1 = 3$ independent degrees of freedom. The antisymmetric
tensor $B_{\mu \nu}$ 
satisfies $\partial_{\mu} B_{\mu \nu} = 0$, which constitute $d - 1  =
3$ constraints in 
$d = 4$. The remaining 3 degrees of freedom in $B_{\mu \nu}$ can be
represented in terms 
of a conserved monopole current density $\widetilde{q}_{\mu}$ (with
$\partial_{\mu} \widetilde{q}_{\mu} = 0$) in the form
\beq
\label{2.5}
{1 \over 2} \varepsilon_{\sigma \mu \nu \rho} \ \partial_{\mu} B_{\nu
\rho} =  \widetilde{q}_{\sigma} \quad .
  \eeq 

\noi Integrating (\ref{2.5}) over a lattice cube with surface $\sum_i$
one obtains 
\beq
\label{2.6}
{1 \over 2} \oint_{\sum_i} B_{\mu \nu} \ d\sigma_{\mu \nu \rho} =
  q_{\rho} (z_i) \quad . 
  \eeq

\noi The integer monopole currents $q_{\rho}$, situated at centres $z_i$
of the lattice cubes, are related to the density
$\widetilde{q}_{\sigma}$ by 
\beq
\label{2.7}
\widetilde{q}_{\rho} (z) = \sum_i q_{\rho} (z_i) \ \widetilde{\delta}(z - z_i) 
  \eeq 

\noi where $\widetilde{\delta}(z - z_i)$ denotes the Kronecker symbol,
$\widetilde{\delta}(z - z_i) = 1$ for  $z = z_i$, $\widetilde{\delta}(z
- z_i) = 0$ otherwise. Introducing a dual field strength $H_{\sigma}$
for $B_{\mu \nu}$,  
\beq
\label{2.8}
H_{\sigma} = {1 \over 2} \ \varepsilon_{\sigma \mu \nu \rho} \
\partial_{\mu} B_{\nu \rho} \quad ,  
\eeq

\noi the sum over $n_{\mu \nu}$ in the partition function (\ref{2.3})
can be replaced by sums over $m_{\mu}$ and $B_{\mu \nu}$, together with
the corresponding constraints: 
\bea
\label{2.9}
&&Z = \sum_{m_{\mu}(x)} \sum_{B_{\mu \nu}(x)} \int_{- \pi}^{+ \pi}
\prod_x \left [ dA_{\mu} \ \delta \left ( \partial_{\mu} A_{\mu}\right )
\ \widehat{\delta} \left ( \partial_{\mu} m_{\mu} \right ) \
\widehat{\delta}^{d-1} \left ( \partial_{\mu} B_{\mu \nu} \right )
\right ] \nn \\  
&&\times \left ( \sum_{\widetilde{q}(z)} \prod_z \widehat{\delta}^{d-1}
\left ( H_{\sigma}(z) - \widetilde{q}_{\sigma}(z) \right ) \right ) \
e^{-{1 \over 4e^2} \sum\limits_x ( F_{\mu \nu} - 2 \pi \partial_{[\mu}
m_{\nu ]} - 2 \pi B_{\mu \nu})^2} 
\quad .
 \eea

\noi Here $\widehat{\delta}$ denote again Kronecker symbols, now in
field space. The $d- 1$ dimensional Kronecker symbol of a conserved
vector $v_{\mu}$, $\widehat{\delta}^{d-1}(v_{\mu})$ with $\partial_{\mu}
v_{\mu} = 0$, can be represented as
\beq
\label{2.10}
\widehat{\delta}^{d-1} (v_{\mu}) = {\rm const.} \int_{-\pi}^{+ \pi}
\prod_x \left [ d C_{\mu} \delta \left ( \partial_{\mu} C_{\mu} \right )
\right ] \ e^{i\sum\limits_x C_{\mu} v_{\mu}} \quad .  \eeq   

\noi Next we can combine the gauge field $A_{\mu}$ with the integers
$m_{\mu}$ into a single field $A'_{\mu}$, which varies from $- \infty$
to $+ \infty $ at each lattice site:
\beq
\label{2.11}
A'_{\mu} = A_{\mu} + 2 \pi m_{\mu} \quad .
\eeq  

\noi At this point the introduction of the gauge fixing $\delta$
function in (\ref{2.3}) proves to be convenient. Omitting the primes of
$A'_{\mu}$, the partition function becomes
\bea
\label{2.12}
&&Z = \sum_{B_{\mu \nu}(x)} \int_{-\infty}^{+ \infty} \prod_x \left [ d
A_{\mu} \delta \left ( \partial_{\mu} A_{\mu} \right ) \
\widehat{\delta}^{d-1} \left ( \partial_{\mu} B_{\mu \nu} \right )
\right ] \left ( \sum_{\widetilde{q}_{\sigma}(z)} \prod_z
\widehat{\delta}^{d-1} \left ( H_{\sigma} (z) -
\widetilde{q}_{\sigma}(z) \right ) \right ) \nn \\
&&\qquad \times e^{-{1 \over 4e^2} \sum\limits_x (F_{\mu \nu} - 2 \pi
B_{\mu \nu})^2} \quad 
. \eea 

Our next aim is the explicit evaluation of the sum over 
monopole current configurations $\widetilde{q}_{\sigma}(z)$. First, we
decompose all possible 
configurations $\widetilde{q}_{\sigma}(z)$ into configurations, which are
nonvanishing at $N$ centres of the lattice cubes:
\beq
\label{2.13}
\left . \sum_{\widetilde{q}_{\sigma}(z)} = \sum_{N = 0}^{\infty} {1
\over N !} \sum_{z_1 \dots z_N} \ \sum_{q_{\sigma}(z_i)\not= 0} \ 
\right |_{\ \widetilde{q}_{\sigma}(z) = \sum\limits_{i=1}^N
q_{\sigma}(z_i)\widetilde{\delta}(z -z_i)} \quad . \eeq    

\noi Second, we observe that contributions to the partition function
with $|q_{\sigma}(z_i)| \geq 2$ are suppressed relatively to
configurations with $q_{\sigma}(z_i) = \pm 1$ \cite{3r}; therefore we
will restrict the sum over $q_{\sigma}(z_i)$ to these values
subsequently. Third, the corresponding term $\sim B_{\mu \nu}^2$ in the
action in (\ref{2.12}) is problematic even for these restricted
configurations: Near a monopole current situated at $z'$ the field
$B_{\mu \nu}(x)$ 
behaves like $|x - z'|^{-3}$ (in $d = 4$), consequently the continuum
integral $\int d^4x B_{\mu \nu}^2$ would diverge quadratically, whereas
on the lattice we are left with an ambiguity in the form of a factor
\beq
\label{2.14}
\xi = e^{- {{\rm const.} \over e^2}}
\eeq  

\noi per centre with nonvanishing monopole current. Taking the $N$
powers of $\xi$ into account we thus rewrite the sum over
$\widetilde{q}_{\sigma}(z)$ as 
\bea
\label{2.15}
&&\sum_{\widetilde{q}_{\sigma} (z)} \prod_z \widehat{\delta}^{d-1} \left
( H_{\sigma}(z) - \widetilde{q}_{\sigma}(z) \right ) \nn \\ 
&& \rightarrow \prod_z \sum_{N=0}^{\infty} {1 \over N !} \ \xi^N
\sum_{z_1 \dots z_N} \ \sum_{q_{\sigma}(z_i) = \pm 1}
\widehat{\delta}^{d-1} \left ( H_{\sigma} (z) - 
\sum_{i=1}^N q_{\sigma} (z_i) \widetilde{\delta}(z - z_i) \right ) \ .
 \eea

\noi Let us now fix $z$, $N$ and the Lorentz index $\sigma$ and
investigate, in which cases the sums over $z_i$ and $q_{\sigma}(z_i)$
give a nonvanishing contribution to the monopole current density
\beq
\label{2.16}
\widetilde{q}_{\sigma}(z) = \sum_{i=1}^N q_{\sigma}(z_i) \
\widetilde{\delta}(z - z_i) 
\eeq 

\noi in the argument of the Kronecker symbol
$\widehat{\delta}^{d-1}(H_{\sigma}(z) - \widetilde{q}_{\sigma}(z))$. Let
us assume that the lattice has $V$ sites. Then, a sum 
over a variable $z_i$ and the associated monopole current
$q_{\sigma}(z_i) = \pm 1$ 
gives $2V$ terms. In $(2V - 2)$ cases the contribution to
$\widetilde{q}_{\sigma}(z)$ 
vanishes, since $z_i$ differs from $z$ and $\widetilde{\delta}(z - z_i)$
is zero. The 
two remaining cases give contributions $\Delta \widetilde{q}_{\sigma} =
\pm 1$ to 
$\widetilde{q}_{\sigma}$. Turning to the sums over $N$ variables $z_i$
and associated 
monopole currents $q_{\sigma}(z_i) = \pm 1$ we can decompose the result
into powers of 
$(2V - 2)$: We have $(2V - 2)^N$ cases, where no $z_i$ coincides with
$z$, and where 
$\widetilde{q}_{\sigma}$ vanishes. We have $2N(2V - 2)^{N-1}$ cases,
where one $z_i$ 
coincides with $z$; these $2N(2V - 2)^{N-1}$ cases can be decomposed
into $N(2V - 2)^{N-1}$ cases with $\widetilde{q}_{\sigma} = + 1$ and
$N(2V - 2)^{N-1}$ cases with $\widetilde{q}_{\sigma} = - 1$. Next we
have ${1 \over 2} \cdot (2N)\cdot (2N - 2)\cdot (2V - 2)^{N-2}$ cases
where two $z_i$ coincide with $z$; a quarter of them corresponds 
to $\widetilde{q}_{\sigma} = + 2$, half of them to
$\widetilde{q}_{\sigma} = 0$ (since 
$\widetilde{q}_{\sigma}(z_i) = - \widetilde{q}_{\sigma}(z_j)$ with $z_i
= z_j = z$), 
and the remaining quarter to $\widetilde{q}_{\sigma} = - 2$. In general
we have ${N 
\choose m} (2V-2)^{N-m} \ 2^m$ cases where $m$ variables $z_i$ coincide
with $z$. The 
corresponding contributions to $\widetilde{q}_{\sigma}$ are generally
different; if 
we distinguish these contributions to $\widetilde{q}_{\sigma}$ we can
write the different cases as 
\beq
\label{2.17}
\left . {N \choose m} (2 V - 2)^{N-m} \left [ \sum_{\nu = 0}^m {m
\choose \nu } \right |_{\widetilde{q}_{\sigma} = 2 \nu - m} \right ]
\eeq 

\noi where the expression in the squared brackets gives $2^m$
terms. Summing over 
$m$ we can thus rewrite the expression (\ref{2.15}) as
\beq
\label{2.18}
\prod_z \sum_{N= \nu}^{\infty} {1 \over N !} \ \xi^N \sum_{m=0}^N {N
\choose m} (2V - 2)^{N-m} \sum_{\nu = 0}^m {m \choose \nu}
\widehat{\delta}^{d-1} \left ( H_{\sigma} (z) - 2 \nu + m \right ) \quad
. \eeq   
      
\noi The sums can be rearranged and partially evaluated; as an
intermediate result one obtains
\beq
\label{2.19}
\prod_z e^{\xi (2V-2)} \sum_{\nu = 0}^{\infty} \sum_{n=0}^{\infty}
\xi^{n+\nu} {1 \over n ! \nu !} \ \widehat{\delta}^{d-1} \left
( H_{\sigma}(z) - \nu + n \right ) \quad . 
\eeq  

\noi Employing the Kronecker symbol in order to perform either of the
sums over $\nu$ or 
$n$ one ends up with a standard series representation for a modified
Bessel function 
\cite{12r} and, omitting the field independent prefactor $\exp \xi (2V -
2)$, the result becomes simply
\beq
\label{2.20}
\prod_{z, \sigma} I_{H_{\sigma}(z)} \ (2 \xi) = e^{\sum\limits_{z,
\sigma} \log I_{H_{\sigma}(z)} (2\xi)}  \eeq  

\noi where we have restored the summation over the Lorentz index
$\sigma$. \par 

The same result has been obtained previously by Diamantini, Quevedo and
Trugenberger 
\cite{5r} with the help of an exact duality transformation of the
partition function of 
the massive dual gauge field. As noted in \cite{5r}, in the
semiclassical approximation 
$H_{\sigma} \to \infty$, $\xi \to \infty$, $H_{\sigma}/\xi$ fixed the
exponent in (\ref{2.20}) becomes \cite{13r}
\beq
\label{2.21}
\sum_z \left ( - H_{\sigma} \ {\rm arcsinh} \left ( {H_{\sigma} \over 2
\xi} \right ) + \sqrt{H_{\sigma}^2 + 4 \xi^2} \right ) \equiv - S_P(H)
\eeq  

\noi whereupon one recovers the Euclidean four-dimensional version of
Polyakov's action $S_P(H)$ in \cite{4r}.  \par

With the result for the summation over monopole configurations at hand
we are in a 
position to rewrite the partition function (\ref{2.12}). For convenience
we switch to a 
continuum notation, and rescale the fields such that the kinetic terms
are properly 
normalized: $A_{\mu} \to e \Lambda_0^{-1} A_{\mu}$, $B_{\mu \nu} \to
\sqrt{2\xi} \Lambda_0^{-1}
B_{\mu \nu}$ such that $\Lambda_0^{4} S_P(H) = {1 \over 2} H_{\mu}^2 +
{\cal O}(H^4)$, 
where $\Lambda_0$ is the inverse lattice spacing. We obtain
\beq
\label{2.22}
Z = \int {\cal D} A_{\mu} {\cal D} B_{\mu \nu} \delta (\partial_{\mu}
A_{\mu}) \delta^{d-1}(\partial_{\mu}B_{\mu \nu}) \ e^{- \int d^4x \{{1
\over 4} (F_{\mu \nu} - m B_{\mu \nu} )^2 + S_P(H)\}} \eeq

\noi where the mass $m$ of the Kalb-Ramond field $B_{\mu \nu}$ is given
by 
\beq
\label{2.23}
m^2 = {8 \pi^2 \over e^2} \ \xi \Lambda_0^2 \quad .
\eeq    

\noi In the weak field limit of $S_P(H)$, the action in (\ref{2.22})
coincides with the 
$4d$ version of the action proposed by Quevedo and Trugenberger
\cite{2r} to describe the condensation of topological defects. \par

Now there are two options concerning a subsequent treatment of the
partition function (\ref{2.22}): The first option consists in a field
redefinition 
\beq
\label{2.24}
B_{\mu \nu} \to B_{\mu \nu} + {1 \over m} \ F_{\mu \nu} \quad .
\eeq 

\noi Since $H_{\mu}$ is invariant under this redefinition of $B_{\mu
\nu}$ by a Bianci 
identity, the gauge field $A_{\mu}$ disappears completely from the
action and appears 
only in the $\delta$ functions $\delta (\partial_{\mu}A_{\mu})
\delta^{d-1} 
(\partial_{\mu} B_{\mu \nu} + {1 \over m} \partial_{\mu} F_{\mu
\nu})$. Since the number 
of $\delta$ functions matches precisely the number of degrees of freedom
of $A_{\mu}$, 
the $A_{\mu}$ path integral can be performed giving just trivial (field
independent) determinants. The resulting partition function reads
\beq
\label{2.25}
Z^{(1)} = \int {\cal D} B_{\mu \nu} \ e^{-\int d^4x\{ {1 \over 4} m^2
B_{\mu \nu}^2 + 
S_P(H)\}} \quad ,  \eeq 

\noi which is the version derived by Polyakov \cite{4r}. \par 

The second way to treat the partition function (\ref{2.22}) consists in
the standard procedure to promote the $\delta$ functions to gauge fixing
terms in the action: First, we replace the $\delta$ functions by
\beq
\label{2.26}
\delta \left ( \partial_{\mu} A_{\mu} - C \right ) \ \delta^{d-1} \left
( \partial_{\mu} B_{\mu \nu} - C_{\nu} \right ) \eeq 

\noi where $C$, $C_{\mu}$ are arbitrary functions with $\partial_{\mu}
C_{\mu} = 0$. 
Next, we integrate over these functions with a Gaussian measure
involving arbitrary 
gauge fixing parameters $\alpha$ and $\beta$; the resulting partition
functions reads (again up to field independent Fadeev-Popov determinants)
\beq
\label{2.27}
Z^{(2)} = \int {\cal D} A_{\mu} {\cal D} B_{\mu \nu} \ e^{-S_{inv} - S_{gf}}
\eeq  

\noi with
\bea
\label{2.28}
&&S_{inv} = \int d^4 x \left \{ {1 \over 4} \left ( F_{\mu \nu} - m
B_{\mu \nu} \right )^2 + S_P(H) \right \} \quad , \nn \\ 
&&S_{gf} = \int d^4x \left \{ {1 \over 2 \alpha} \left ( \partial_{\mu}
A_{\mu} \right ) + {1 \over 2 \beta} \left ( \partial_{\mu} B_{\mu \nu}
\right )^2 \right \} \quad . 
 \eea

\noi Here $S_{gf}$ serves to ``gauge fix'' the gauge symmetries of
$S_{inv}$:  
\bea
\label{2.29}
&&{\rm a)} \quad \delta A_{\mu} = \partial_{\mu} \Lambda \quad , \qquad
\delta B_{\mu \nu} = 0  \nn \\ 
&&{\rm b)} \quad \delta A_{\mu} = m \Lambda_{\mu}  \quad , \qquad \delta
B_{\mu \nu} = \partial_{\mu} \Lambda_{\nu} - \partial_{\nu}
\Lambda_{\mu} \quad . 
 \eea

\noi These gauge symmetries are actually not independent; the
transformation a) is 
obtained from b) after identifying $\Lambda_{\mu} = m^{-1}
\partial_{\mu} \Lambda$. A 
remnant of the gauge invariance b) is the following Ward identity, which
is satisfied by the total action $S_T = S_{inv} + S_{gf}$:
\beq
\label{2.30}
{\delta S_T \over \delta A_{\mu}} + {2 \over m} \ \partial_{\nu}
{\delta S_T \over \delta B_{\mu \nu}} + {1 \over \alpha} \
\partial_{\mu} \partial_{\nu} A_{\nu} - {\sq \over \beta m} \
\partial_{\nu} B_{\nu \mu} = 0 \quad . \eeq   

\noi The standard Ward identity related to the gauge invariance a) is
obtained by contracting (\ref{2.30}) with $\partial_{\mu}$. \par

Herewith we conclude this section and turn now to a possible relation
between $S_T$ and the low energy effective action of pure Yang-Mills
theories. 

\mysection{Wilsonian RG flow for Yang-Mills theories with antisymmetric 
tensor fields} 
\hspace*{\parindent} Let us start this section with the definition of 
the partition function of an Euclidean Yang-Mills theory. Subsequently we
will employ the maximal abelian gauge [7,9,14]. Abelian gauges were
originally introduced by t'Hooft [1], who showed that they lead to the
appearance of magnetic monopoles in Yang-Mills theories. Below it will be
useful that abelian Ward identities related to U(1) subgroups of SU(N)
remain valid in the maximal abelian gauge [14]. Including gauge fixing 
terms and ghosts, the Yang-Mills partition function reads:

\beq
\label{3.1}
\exp \left \{ - G(J, \chi, \bar{\chi}) \right \} = 
\int {\cal D}_{reg} (A, c, \bar{c}) \exp \Big \{ - S_{YM} - S_{gf} - 
S_{gh} + J \cdot A + \bar{\chi} \cdot c + \chi \cdot \bar{c} \Big \}  
\eeq

\noi where we used the short-hand notation

\beq
\label{3.2}
J\cdot A = \int d^4x \ J_{\mu}^{\alpha} (x) \ 
A_{\mu}^{\alpha} (x) \qquad \hbox{etc.} \eeq

\noi The index ``reg'' attached to the path integral measure indicates 
an ultraviolet regularisation. $S_{YM}$ denotes the standard Yang-Mills 
action, $S_{gf}$ the gauge fixing terms, and $S_{gh}$ the terms 
depending on the ghost fields. In the maximal abelian gauge it is 
convenient to adopt following conventions: We decompose the
$N^2-1$ generators of SU(N) indexed by $\alpha, \beta = 1 \dots N^2 - 
1$ into the $N-1$ generators of the $N-1$ U(1) subgroups with $N-1$ 
indices $a$, $b$ and the $N(N-1)$ non-diagonal ``charged'' generators 
indexed by $i,j = 1 \dots N(N-1)$. It is helpful to introduce a 
U(1)-covariant derivative $D_{\mu}$, which acts on the charged
fields $\varphi^i = \{A_{\mu}^i , c^i, \bar{c}^i \}$ as
   
\beq
\label{3.3}
D_{\mu} \varphi^i = \partial_{\mu} \varphi ^i + g f^i _{aj} A_{\mu}^a 
\varphi^j \quad . \eeq

\noi The Yang-Mills action thus decomposes as

\beq
\label{3.4}
S_{YM} = \int d^4x {1 \over 4} F_{\mu \nu}^{\alpha} F_{\mu \nu}^{\alpha} 
= \int d^4 x \left \{ {1 \over 4} F_{\mu \nu}^i F_{\mu \nu}^i + 
{1 \over 4} F_{\mu \nu}^a F_{\mu \nu}^a \right \} \quad . \eeq

\noi The maximal abelian gauge corresponds to gauge fixing terms of the 
form

\beq
\label{3.5}
S_{gf} = \int d^4 x \left \{ {1 \over {2 \alpha}} (\partial_\mu A_\mu^a)^2
+ {1 \over {2 \alpha^{(c)}}} (D_\mu A_\mu^i)^2 \right \}
\quad , \eeq

\noi i.e. the gauge fixing of the charged gauge fields $A_{\mu}^i$ is 
U(1) gauge invariant. The form of $S_{gh}$ is not relevant subsequently.

Now we add collective fields $B^a_{\mu \nu}$ for the U(1) field
strengths $F^a_{\mu \nu}$ to the partition function (3.1).
The addition of collective fields corresponds to a
multiplication of the integrand of the path integral with

\beq
\label{3.6}
1 = {1 \over {\cal N}} \int {\cal D} B \exp \{ - \widehat{S} \} 
\quad , \qquad 
\widehat{S} = \int d^4x \left \{ {1 \over 4} \left ( F^a_{\mu \nu} - 
B^a_{\mu \nu} \right )^2 \right \}  
\eeq

\noi Moreover we add the following source term to the exponent under the path
integral: 

\beq
\label{3.7}
\widehat{J}^a_{\mu \nu} \cdot B^a_{\mu \nu} 
\eeq

\noi If one performs the Gaussian path integral over $B$ in
the presence of the source terms (\ref{3.7}), one finds that the sources
$\widehat{J}^a_{\mu \nu}$ couples to the operator
$F_{\mu \nu}^a$. In addition one obtains terms quadratic in the sources. 
The expression for the partition function finally becomes

\beq
\label{3.8}
\exp \left \{ - G \left (J, \chi , \bar{\chi}, \widehat{J} \right ) \right \} 
= \int {\cal D} \left ( A, c, \bar{c}, B \right )
\exp \left \{ - S(A,B) - S_{gf} - S_{gh} +  Sources \right \}  \eeq

\noi with 

\bea
\label{3.9}
S(A,B) = \int d^4x \left \{ {1 \over 4} F^i_{\mu \nu} F^i_{\mu \nu} +
{1 \over 2} F^a_{\mu \nu} F^a_{\mu \nu} -
{1 \over 2} F^a_{\mu \nu} B^a_{\mu \nu} +
{1 \over 4} B^a_{\mu \nu} B^a_{\mu \nu} \right \} \eea

\noi The expression $Sources$ reads

\beq
\label{3.10}
Sources  = {J} \cdot A + \bar{\chi} \cdot c + \chi \cdot \bar{c}
 + \widehat{{J}} \cdot B    \eeq
 
\noi where we employ the convention (3.2). For later use we introduce
the effective action $\Gamma (A, c, \bar{c}, B)$, the Legendre 
transform of $G(J, \chi, \bar{\chi},\widehat{J})$:
\beq
\label{3.11}
\Gamma (A, c, \bar{c}, B) = G(J, \chi, \bar{\chi},\widehat{J})
+ {J} \cdot A + \bar{\chi} \cdot c + \chi \cdot \bar{c} + \widehat{J} 
\cdot B  \quad .
 \eeq

The Wilsonian ERGEs \citm{15r}{18r} are obtained by adding an
``artifical'' infrared cutoff $k$ to the partition function (\ref{3.1}) 
or (\ref{3.8}). One exploits the facts that the corresponding $k$ 
dependent effective action $\Gamma_k$ a) is equal to the full quantum 
effective action $\Gamma$ for $k = 0$, b) corresponds to
the classical action $S_{cl}$ in the limit $k \to \infty$ (up to 
additional terms determined by modified Slavnov-Taylor identities 
\citd{16r}{17r}), and c) that an exact functional differential equation 
with respect to $k$ (the ERGE) can be derived. The integration of the 
ERGEs from some large value $k = \Lambda$ down to $k = 0$ provides us 
with a non-perturbative method for calculating $\Gamma_{k=0}$ in terms
of some ``high energy'' efective action $\Gamma_{\Lambda} \sim S_{cl}$. 
The formalism can straightforwardly be extended towards partition 
functions involving sources for collective fields \citt{18r}{6r}{11r}, 
provided $S_{cl}$ is replaced by $S_{YM} + \widehat{S} = S(A,B)$. \par

Let us consider directly the partition function (\ref{3.8}) with the 
collective fields included. In the presence of an infrared cutoff it 
becomes

 \bea
\label{3.12}
&&\exp \left \{ - G_k \left (J, \chi , \bar{\chi}, 
\widehat{J} \right ) \right \} \nn \\
&&= \int {\cal D} \left ( A, c, \bar{c}, B \right ) \exp \left \{ -
S(A,B) - S_{gf} - S_{gh} - \Delta S_k  + \ Sources \right \} \nn \\
\eea
  
\noi where $\Delta S_k$ is quadratic in the gluon and ghost fields: 

\beq
\label{3.13}
\Delta S_k = \int {d^4p \over (2 \pi)^4} \Big [ {1 \over 2} 
A_{\mu}^{\alpha}(-p)\  R_{\mu \nu}^{k}(p^2) \ A_{\nu}^{\alpha}(p)  + 
\bar{c}^{\alpha}(-p) \ R_g^{k} (p^2) \ c^{\alpha}(p) \Big ] \quad . \eeq

\noi The functions $R^k(p^2)$ modify the propagators such that modes
with $p^2 \ll k^2$ are suppressed. Possible choices are 

\bea
\label{3.14}
&&R_{\mu \nu}^k(p^2) = \left ( p^2 
\delta_{\mu \nu} - ({1 \over \alpha} - 1) p_\mu p_{\nu} \right ) 
\widetilde{R}^k(p^2) \quad , \nn \\ 
&&R_g^k(p^2) = \ p^2 \ \widetilde{R}^k(p^2) \quad , \nn \\
&&\widetilde{R}^k(p^2) = {e^{-p^2/k^2} \over 1 - e^{-p^2/k^2}} \quad .
\eea

Infrared cutoffs for the collective fields could also be introduced 
\cite{18r} but are not mandatory. The ERGEs for the functional 
$G_k$ follow after differentiation of both sides of (\ref{3.12}) 
with respect to $k$, and expressing the expectation value 
$<\partial_k \Delta S_k>$ through variations with respect to the 
sources. One obtains

\bea
\label{3.15}
\partial_k G_k &=& \int {d^4p\over (2 \pi )^4} \left \{ 
{1 \over 2} \partial_k R_{\mu \nu}^k 
\left ( {\delta G_k \over \delta {J}_{\mu}^{\alpha}(p)} 
{\delta G_k \over \delta {J}_{\nu}^{\alpha}(-p)} - {\delta^2 
G_k \over \delta {J}_{\mu}^{\alpha}(p) \ \delta {J}_{\nu}^{\alpha} (-p)}
\right ) \right . \nn \\ 
&&\left . + \partial_k R_g^k
\left ( {\delta G_k \over \delta \chi^{\alpha}(p)} 
{\delta G_k \over \delta {\bar \chi}^{\alpha}(-p)} - 
{\delta^2 G_k \over \delta \chi^{\alpha} (p) 
\ \delta {\bar \chi}^{\alpha}(-p)} 
\right ) \right \}  \quad .
\eea  

\noi The ERGEs for $\Gamma_k$, the Legendre transform of $G_k$, 
can easily be obtained from (\ref{3.15}), but they are not needed in 
the following. \par

Now we wish to show that a slight variant of the Ward identity (2.30),
which implies the (gauge fixed) vector like gauge symmetry b) in 
eq. (2.29),
constitutes a (quasi) fixed point of the ERGEs. To this end we study the
following functional ${\Omega}^{k,a}_{\mu}$, which is constructed in
terms of sources, $G_k$ (satisfying the ERGEs (3.15)), and the IR cutoff
function $R^k$:

\beq
\label{3.16} 
\Omega_{\mu}^{k,a} = J_{\mu}^a (p) + {2i \over m} p_{\nu}
\widehat{J}_{\mu \nu}^a(p) + {1 \over \alpha} p_{\mu}p_{\nu}
{\delta {G}_k \over \delta {J}_{\nu}^a(-p)} - {ip^2 \over
\beta m} p_{\nu} {\delta {G}_k \over \delta \widehat{J}_{\nu
\mu}^a (-p)} + R^{k}_{\mu \nu} {\delta {G}_k \over \delta 
{J}_{\nu}^a(-p)} \ . \eeq

\noi First we note that, for $k \to 0$, $R^k$ and hence the last term in
$\Omega_{\mu}^{k,a}$ vanishes. The remaining 4 terms in
$\Omega_{\mu}^{0,a}$ can be 
expressed in terms of $\Gamma_0$, the effective action at vanishing IR
cutoff $k$, 
through the Legendre transform (3.11):

\beq
\label{3.17}
\Omega_{\mu}^{0,a} = {\delta \Gamma_0 \over \delta A_{\mu}^a (-p)} +
{2i \over m} 
p_{\nu} {\delta \Gamma_0 \over \delta B_{\mu \nu}^a(-p)} - {1 \over
\alpha} p_{\mu} 
p_{\nu} A_{\nu}^a (p) + {ip^2 \over \beta m} p_{\nu} B_{\nu \mu}^a(p) 
\ .  \eeq
\noi Thus the functional equation
\beq
\label{3.18}
\Omega_{\mu}^{0,a} = 0 
\eeq

\noi is equivalent to the statement that $\Gamma_0$ satisfies the Ward
identity (\ref{2.30}). \par

Next we consider the variation of $\Omega_{\mu}^{k,a}$ with respect to
the infrared 
cutoff $k$. When one evaluates $\partial_k \Omega_{\mu}^{k,a}$ with 
$\Omega_{\mu}^{k,a}$ as in eq.(3.16), the derivative 
$\partial_k$ hits ${G}_k$ and the IR cutoff function $R^k$. Using
the ERGE (3.15) 
for $\partial_k {G}_k$ one obtains the following important result: 
\beq
\label{3.19}
\partial_k \Omega_{\mu}^{k,a} = 0 \qquad \hbox{if} \quad
\Omega_{\mu}^{k,a} = 0 
\quad . \eeq

\noi Hence the functional equation $\Omega_{\mu}^{k,a} = 0$ -- either in
terms of ${G}_k$ or in terms of $\Gamma_k$ -- constitutes a
quasi-fixed point of the ERGEs. If it 
is satisfied by ${G}_k$ or $\Gamma_k$ for some $k$, it will also be
satisfied by 
${G}_0$ and $\Gamma_0$, i.e. $\Gamma_0$ will satisfy the Ward
identity (\ref{2.30}). \par

Let us define an ``Abelian projection'' $\bar{\Gamma}_0$ of $\Gamma_0$
by 
\beq
\label{3.20}
\left . \bar{\Gamma}_0 = \Gamma_0 \right |_{\varphi^i = 0} \quad ,
\eeq 

\noi where $\varphi^i$ denote all ``charged'' fields $A_{\mu}^i$, $c^i$,
$\bar{c}^i$ 
with respect to the U(1) subgroups, cf. our convention for the indices
below eq. (3.2). 
Trivially, once $\Gamma_0$ satisfies the Ward identity (\ref{2.30}), it
is also 
satisfied by $\bar{\Gamma}_0(A_{\mu}^a, B_{\mu \nu}^a)$. Up to terms
quadratic in the 
fields and derivatives, the general solution for $\bar{\Gamma}_0$ is
then necessarily 
of the form of the weak field limit of $S_{inv} + S_{gf}$ in
eqs. (\ref{2.28}), i.e. 
of the form of the Quevedo-Trugenberger action describing the
condensation of magnetic 
monopoles. The terms involving higher powers of the dual field strength
$H_{\mu}$ of 
the Kalb-Ramond field $B_{\mu \nu}$, which appear in Polyakov's action
$S_P(H)$ in 
(\ref{2.21}) and (\ref{2.28}), also satisfy the Ward identity, but
cannot be derived by the Ward identity alone. \par

Remember that, in the process of computing $\Gamma_{k=0}$ by integrating
the ERGEs with 
respect to $k$, a ``boundary condition'' $\Gamma_{\Lambda}$ at some
large scale $k = 
\Lambda$ has to be specified. In our case $\Gamma_{\Lambda}$ is given by
the action 
$S(A,B)$ of eq. (3.9), up to gauge fixing terms for the gluons, and up
to additional 
terms of ${\cal O}(g^2R^{\Lambda})$ in order to satisfy the modified
Slavnov-Taylor 
identities [16,17]. It is easily checked that the Abelian projection
$\bar{\Gamma}_{\Lambda}$ does not satisfy the Ward identity (2.30) (for
$\beta \to \infty$), since the three terms ${1 \over 2} F^a F^a - {1
\over 2} F^a B^a + {1 \over 4} B^a B^a$ are not of the form of a
square. Trivially, the classical Yang-Mills action 
does not describe confinement by monopole condensation, even if
collective fields $B_{\mu \nu}^a$ are introduced. \par

During the ERGE flow all parameters in $\Gamma_k$ will vary with $k$;
the ERGE flow can 
be represented as a motion in the infinite dimensional space of
couplings (parameters) 
of $\Gamma_k$. Couplings, which are absent in $\Gamma_{\Lambda}$, but
not protected by 
the modified Slavnov-Taylor identities, will become non-zero, as powers
of $H_{\mu}$ or 
terms of the form $(\partial_{\mu}B_{\mu \nu})^2$. A priori it is an
open question 
whether, for some value of $k$, the terms involving $F^a$ combine with
terms involving 
$B^a$ into combinations of the form $(F^a - m B^a)$, where a
$k$-independent parameter 
$m$ is generated dynamically by dimensional transmutation. A necessary
condition for 
this to happen is that the ``fixed point'' $\Omega_{\mu}^{k,a} = 0$,
with some 
arbitrary ($k$ independent dynamically generated value) of $\beta$, is
infrared attractive.   

Indications for such an ``infrared attractiveness'' can be obtained from
the results in \cite{6r}, where the RG flow of the $A_{\mu}/B_{\mu\nu}$ 
system has been studied in a simple ap\-pro\-xi\-ma\-tion (However, in 
\cite{6r} antisymmetric tensor fields were introduced for all $N^2-1$ 
components of $F_{\mu \nu}$, and the Landau gauge was employed): 
Within a parametrization of $\Gamma_k (A, B)$ of the form

\beq
\label{3.21}
\Gamma_k (A, B) = {Z \over 4} (F_{\mu \nu})^2 - {n \over 2} F_{\mu \nu} 
B_{\mu \nu} + {m^2 \over 4} (B_{\mu \nu})^2 + \hbox{gauge fixing terms} 
\eeq

\noi it was shown that

\beq
\label{3.22}
Z_{eff} = Z - {n^2 \over m^2} = 0 \qquad ,
\eeq
  
\noi whereupon $F$ and $B$ combine into a perfect square, constitutes 
an infrared fixed point. At the starting point, where
$\Gamma_{\Lambda}(A,B)$ is given by $S(A,B)$ of eq. (3.9), we have
$Z=2$, $n = m$ and hence $Z_{eff}(\Lambda ) = 1$. However, already 
perturbatively, $Z_{eff}(k)$ decreases with decreasing $k$ and
approaches thus the fixed point (\ref{3.22}). \par

Within the simple approximation in \cite{6r}, on the other hand, a 
Landau singularity in the running gauge coupling prevented a detailed 
analysis of the regime $k \to 0$. This problem disappeared within a 
less trivial truncation of $\Gamma_k$ in \cite{11r},
where the gauge coupling became even vanishingly small for $k \to 0$. 
\par

Whereas the dependence of the results of \citd{6r}{11r} on the 
truncation of $\Gamma_k$ is an open problem, we emphasize that the 
fixed point nature of the Ward identity, eq. (3.19), is completely 
general. \par

A final remark concerns the relevance of the maximal abelian gauge for
our results: Since the abelian Ward identities (several in the case of
several U(1)'s) follow from the vector Ward identity ${
\Omega}_{\mu}^{k,a} = 0$ after contraction with $p_{\mu}$, their
validity is a necessary condition on $\Gamma_k$, if $\Gamma_k$ is
assumed to satisfy ${\Omega}_{\mu}^{k,a} = 0$. This necessary
condition is guaranteed to be satisfied precisely in the maximal 
abelian gauge. 

\mysection{Summary and Conclusions}
\hspace*{\parindent} The aim of the present paper is to emphasize the
role of 
antisymmetric tensor fields for the description of monopole condensation
in strongly 
coupled gauge theories. In the first part of the paper we have studied
four-dimensional 
compact QED on the lattice, and we have rederived two equivalent
versions of the 
partition function: The first version involves just a massive
Kalb-Ramond field 
$B_{\mu \nu}$, and the original Abelian gauge field $A_{\mu}$, has
disappeared 
completely (it has been ``eaten'' by $B_{\mu \nu}$ in order to become
massive, in the 
same way as Goldstone bosons are eaten by massive vector fields in the
case of 
spontaneous gauge symmetry breaking). The corresponding action had been
obtained by 
Polyakov \cite{4r} and Diamantini, Quevedo and Trugenberger \cite{5r}
before, 
starting with the dual action involving a massive vector field. Here we
have shown how to obtain this action directly from compact QED. \par

The second version of the partition function involves both the gauge
field $A_{\mu}$ 
and the Kalb-Ramond field $B_{\mu \nu}$, but additional gauge fixing
terms which 
fix, in particular, the vector-like gauge symmetry under which
$B_{\mu\nu}$  
transforms. The quadratic part of the action is a special $(4d)$ case of
actions 
proposed by Quevedo and Trugenberger \cite{2r} in order to describe the
condensation of topological defects. Here we have emphasized a Ward
identity 
related to the vector gauge symmetry, whose validity is a sufficient
condition on the action in order to be of the Quevedo-Trugenberger
form. \par 

In the second part of the paper we have studied the Wilsonian exact 
re\-nor\-ma\-li\-za\-tion group flow of pure Yang-Mills theories in the
maximal Abelian 
gauge, and in the presence of auxiliary fields $B_{\mu \nu}^a$ for the
``diagonal'' 
components $F_{\mu \nu}^a$ of the field strength. We have introduced a
modified 
($k$-dependent) vector Ward identity $\Omega_{\mu}^{k,a}(\Gamma_k) = 0$
and shown that 
its validity is stable under the ERG flow. At va\-ni\-shing IR cutoff
$k$ it coincides 
with the Ward identity above. Its validity does not fix $\Gamma_k$
completely, but 
constrains the infinitely many couplings in $\Gamma_k$ to lay inside a
fixed 
``hyperplane'' in the infinite dimensional space of couplings. This
picture can be 
represented schematically as in Fig.~1: The plane in Fig.~1 represents
the infinite 
dimensional space of couplings of actions depending on Abelian gauge
fields $A_{\mu}$ and 
Kalb-Ramond fields $B_{\mu \nu}$. The curve $W$ represents the
``hyperplane'' on which the vector Ward identity is satisfied. \par

Wilsonian actions of Yang-Mills theories can be represented on this
plane, once 
auxiliary fields $B_{\mu \nu}^a$ are introduced, and once they are
projected onto the 
Abelian subsector. Their ERG flow is represented by the curve $YM$ in
Fig.~1. The 
starting point of the ERG flow is denoted by the point $P$
(``perturbation theory''), 
which is certainly not on the curve $W$. At the point $Q$ the Wilsonian
Yang-Mills 
action would satisfy the vector Ward identity, and the interesting
question is whether 
it is assumed in the limit of vanishing IR cutoff $k$. We have shown
that it is a fixed 
point of the ERG flow, and that it is IR stable in a particular
direction in the space 
of couplings; the general IR stability remains to be
shown. Furthermore, 
perturbatively the ERG flow is from $P$ towards $Q$, which is indicated
by the arrow -- 
this is related to the decrease of the wave function renormalization of
the diagonal 
gluons, which gives the increase of the gauge coupling in the maximal
Abelian gauge due 
to the Abelian Ward identity \cite{14r}. Clearly further investigations
of properties 
of Wilsonian Yang-Mills actions near the point $Q$ are highly desirable.

\newpage
\def\labelenumi{[\arabic{enumi}]}
\noindent
{\Large\bf References}
\ben
\item\label{1r} S. Mandelstam, Phys. Rep. {\bf 23C} (1976) 245; \\
G. t'Hooft, Nucl. Phys. {\bf B190} (1981) 455.  
\item\label{2r} F. Quevedo, C. Trugenberger, Nucl. Phys. {\bf B501}
(1997) 143.  
\item\label{3r} A. M. Polyakov, {\it Gauge Fields and Strings} (Harwood
Academic Publishers, Chur, 1987), chapt. 4.  
\item\label{4r} A. M. Polyakov, Nucl. Phys. {\bf B486} (1997) 23. 
\item\label{5r} M. Diamantini, F. Quevedo, C. Trugenberger,
Phys. Lett. {\bf B396} (1997) 115.  
\item\label{6r} U. Ellwanger, Nucl. Phys. {\bf B531} (1998) 593.
\item\label{7r} K.-I. Kondo, Phys. Rev. {\bf D57} (1998) 7467;
ibid. {\bf D58} (1998) 105016.      
\item\label{8r} M. Halpern, Phys. Rev. {\bf D16} (1977) 1798; \\
M. Schaden, H. Reinhardt, P. Amundsen, M. Lavelle, Nucl. Phys. {\bf
B339} (1990) 595; \\
K. Langfeld, H. Reinhardt, Nucl. Phys. {\bf A579} (1994) 472.
\item\label{9r} M. Quandt, H. Reinhardt, Int. J. Mod. Phys. {\bf A13}
(1998) 4049.  
\item\label{10r} D. Antonov, Mod. Phys. Lett. {\bf A13} (1998) 581,
ibid. 659, Phys. Lett. {\bf B427} (1998) 274; \\
D. Antonov, D. Ebert, Phys. Lett. {\bf B444} (1998) 208.  
\item\label{11r} U. Ellwanger, Eur. Phys. J. {\bf C7} (1999) 673.  
\item\label{12r} I. Gradstheyn and I. Ryzhik, {\it Table of Integrals,
Series, and Products} (Academic Press, Boston, 1980).   
\item\label{13r} Erd\'elyi et al., {\it Higher Transcendental
Functions}, Vol. 2 (Mc Graw Hill, New York, 1953).      
\item\label{14r} M. Quandt, H. Reinhardt, Phys. Lett. {\bf 424} (1998)
115. 
\item\label{15r} K. Wilson, I. Kogut, Phys. Rep. {\bf 12} (1974) 15; \\
F. Wegner, in: Phase Transitions and Critical Phenomena, Vol. 6, eds. 
C. Domb, M. Green (Academic Press, NY 1975); \\
J. Polchinski, Nucl. Phys. {\bf B231} (1984) 269; \\
C. Wetterich, Phys. Lett. {\bf B301} (1993) 90; \\
T. Morris, Int. J. Mod. Phys. {\bf A9} (1994) 2411.    
\item\label{16r} C. Becchi, in: Elementary Particles, Field Theory and 
Statistical Mechanics, eds. M. Bonini, G. Marchesini, E. Onofri, Parma 
University 1993; \\
M. Bonini, M. D'Attanasio, G. Marchenisini, Nucl. Phys. {\bf B418} 
(1994) 81, ibid. {\bf B421} (1994) 429, {\bf B437} (1995) 163, Phys. 
Lett. {\bf B346} (1995) 87.
\item\label{17r} U. Ellwanger, Phys. Lett. {\bf B335} (1994) 364.  
\item\label{18r} U. Ellwanger, C. Wetterich, Nucl. Phys. {\bf B423} 
(1994) 137.{\bf }

\een

\vspace{3cm}
\noi{\Large\bf Figure Caption} \par \vskip 5 truemm
\begin{description}
\item{\bf Fig. 1:}  Schematic representation of the space of couplings 
of effective actions depending on abelian gauge fields $A_{\mu}$ and 
antisymmetric tensor fields $B_{\mu\nu}$. The meaning of the curves
$W$, $YM$ and the points $P$, $Q$ is explained in section~4. 
\end{description}

\newpage
\begin{figure}[p]
\unitlength1cm
\begin{picture}(1,1)
\put(0.0,-2.0){
\epsffile{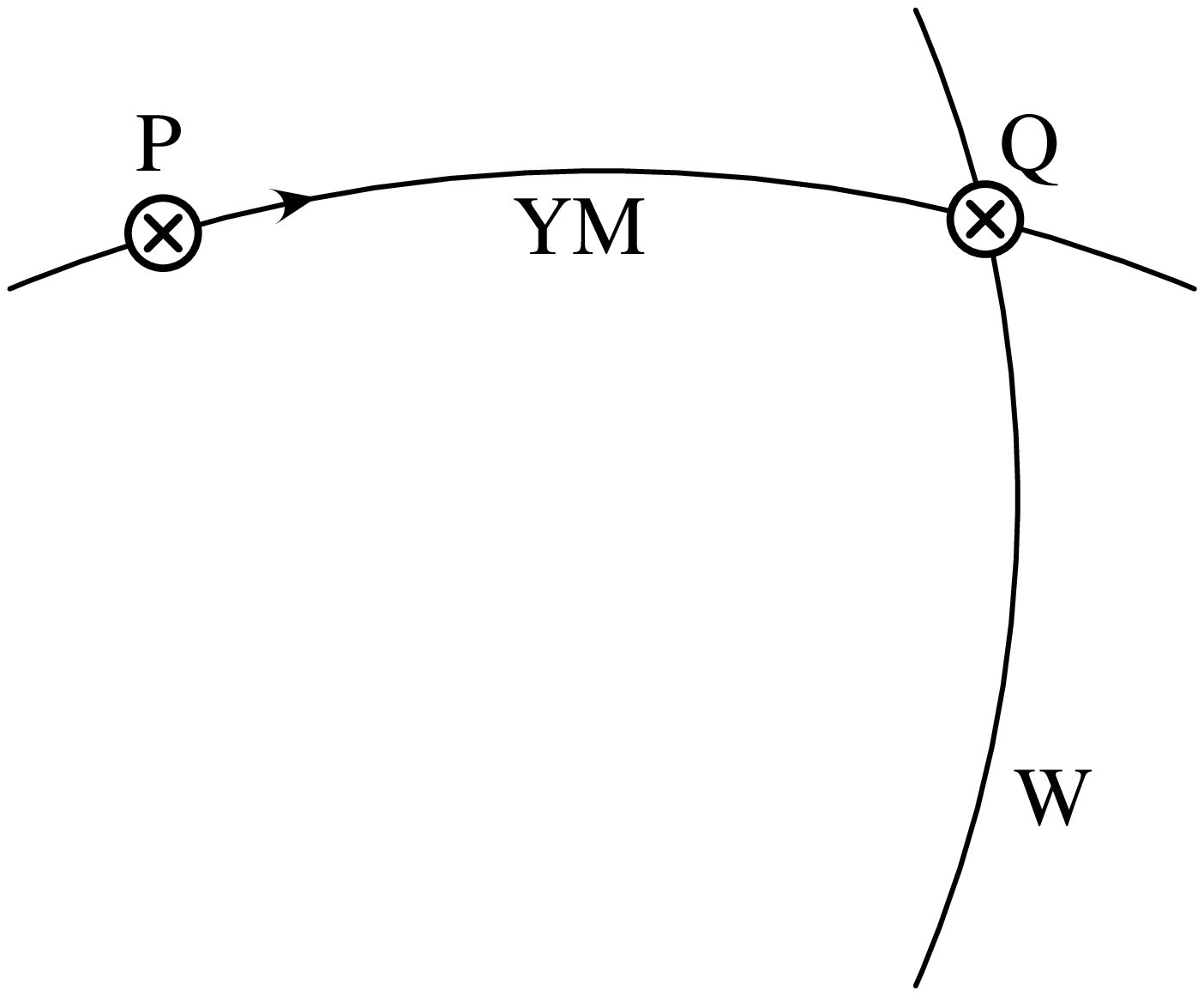}}
\put(7.0,-8) {\bf Fig. 1}
\end{picture}\par
\end{figure}

\end{document}